\documentclass[12pt]{article}
\def\endnote{\footnote}
\begin{document}
\begin{center}
\large\textbf{Philosophical Implications of Inflationary
Cosmology}
\end{center}

\begin{center}
Joshua Knobe,
Ken D. Olum, and Alexander Vilenkin
\end{center}

\begin{abstract}
Recent developments in cosmology indicate that every history having a 
nonzero probability is realized in infinitely many distinct regions of 
spacetime. Thus, it appears that the universe contains infinitely many 
civilizations exactly like our own, as well as infinitely many civilizations 
that differ from our own in any way permitted by physical laws. We explore 
the implications of this conclusion for ethical theory and for the doomsday 
argument. In the infinite universe, we find that the doomsday argument 
applies only to effects which change the average lifetime of all 
civilizations, and not those which affect our civilization alone.

\tableofcontents
\end{abstract}

\section{Introduction}

It is said that the ancient Greek philosopher Diodorus Cronos once put forth 
a powerful argument for a peculiar view about the relationship between the 
possible and the actual. Diodorus claimed that everything that could 
possibly happen is either occurring right now or will occur at some point in 
the future. His claim, in other words, was that there are no unrealized 
possibilities. Unfortunately, the works of Diodorus have been lost, and 
although a number of modern philosophers have tried valiantly to reconstruct 
his argument, no one really knows exactly how it was supposed to go. 

Nonetheless, we think that Diodorus's conclusion was essentially correct, 
and we will here provide a new, entirely modern argument for it. Unlike the 
original argument of Diodorus, however, our argument draws on inflationary 
cosmology and quantum mechanics. It follows from inflationary cosmology that 
the universe is infinite and can therefore be divided into an infinite 
number of regions of any given size. But it follows from quantum theory that 
the total number of histories that can occur in any one of these regions in 
a finite time is finite. We draw on these two premises to argue for our 
central conclusion: that all possible histories are realized in some region 
of the universe.

This conclusion must be understood in a special sense. First, note 
that we are reserving the word `possible' for \textit{physical 
}possibilities. Thus, although it might be metaphysically possible 
for a system to violate strict physical laws, such a system would not count 
as `possible' in our sense. Second, we do not mean to say that anything 
that can possibly happen to a token individual will actually happen to that 
individual. The conclusion is rather that anything that can happen to a 
particular type of individual will actually happen to some token of that 
type. So, for example, we do not claim that everything that can possibly 
happen to you will actually happen to \textit{you}. The 
claim is only that everything that can possibly happen to you will actually 
happen to some qualitative duplicate of you. 

Ultimately, our argument is more a scientific theory than a philosophical 
account, and it has already been presented as such elsewhere (Garriga {\&} 
Vilenkin [2001]). Still, we feel that the theory has important implications 
for issues that have traditionally been the concern of philosophers. This 
paper, written jointly by two physicists and a philosopher, explores these 
philosophical implications.

We proceed in two steps. First, we provide a condensed, non-technical 
explanation of the argument. Then we explore the implications of this 
argument for questions about modality, ethics, and doomsday. 

\section{Physics Background}

The assertions that the universe is infinite and that the number of possible 
histories in a finite spacetime region is finite are crucial for our 
argument. Here, we shall briefly discuss the physical origin of these claims 
and provide some references where further details can be found. 

\subsection{The number of possible histories is finite}

Suppose we pick a region of space and an interval of time. This defines a 
region of spacetime. We want to consider histories that can occur in this 
spacetime region. If we divide the space in such a region into small 
subregions, we can define a history as a specification of the contents of 
each subregion at successive moments of time.

Quantum mechanics assigns a probability to each of the histories, and we say 
that a history is possible if its probability is not equal to zero. This 
includes a very wide class of histories, since in quantum mechanics anything 
that is not strictly forbidden has a nonzero probability. The only histories 
that are excluded are the ones that violate some exact conservation laws, 
like the conservation of energy or of electric charge.

It can be shown, however, that there are only finitely many distinct 
histories that can occur in any finite spacetime region. One might think 
that the subregions and the intervals between moments of time could be made 
arbitrarily small, and the contents specified arbitrarily precisely, so the 
number of possibilities should be infinite, and of course in classical
mechanics that would indeed be so.  But in the quantum mechanical world
the situation is different.

In quantum mechanics, if two histories are too similar, there is the
possibility of interference between them.  In that case, it is not
meaningful to say that the two histories are alternative
possibilities.  Instead, both possibilities together contribute to the
outcome.

The concept in quantum mechanics that corresponds to the ordinary idea
of alternative possibilities is that of decoherent histories
(Gell-Mann and Hartle, [1993]).  If the subregions and their values are
specified sufficiently coarsely, the resulting histories will
decohere, meaning that they do not interfere with each other and can
be meaningfully interpreted as classical alternatives.  When we
discuss the set of possible histories, we mean the set of decoherent
histories, which are mutually exclusive possibilities for the
evolution of a region.  In order to have decoherence between the
histories, the subregions cannot be too small and their values cannot
be too finely specified, and as a result the number of decoherent
histories is finite.  For more details see Garriga and Vilenkin
([2001]).

Suppose, for example, that we have a box containing a radioactive 
atom, which decays by emitting an alpha particle. We observe the box over a 
finite time T. Histories of the box can be defined by dividing it into 
little cells and subdividing T into small intervals. If the decay occurs 
during the time T, then at some point an alpha-particle will appear in one 
of the cells, and then move on to other cells, as it propagates away from 
the atom. A possible history is given by specifying the cells containing the 
atom and the alpha-particle at each time interval. To assure decoherence, 
the cell sizes need to be chosen greater than the de Broglie wavelength of 
the alpha-particle; there is a similar lower bound on the time 
interval.

\subsection{The universe is infinite}

The claim that the universe is infinite is a consequence of the theory of 
inflation. This theory began as a speculative hypothesis when it was 
proposed by Alan Guth ([1981]), but it is now well on its way to becoming one 
of the cornerstones of modern cosmology. The central role in the theory is 
played by a peculiar form of matter---known as `false vacuum'---which 
is characterized by high energy and strong repulsive gravitational field. 
Here, the word `false' alludes to the fact that this type of vacuum is 
unstable and decays into ordinary (true) vacuum. Inflation is an epoch of 
super-fast, accelerated cosmic expansion, driven by the repulsive gravity of 
false vacuum. Decay of the false vacuum marks the end of inflation and plays 
the role of the big bang in this theory.

One of the striking aspects of inflation is that, generically, it never ends 
in the entire universe. False vacuum decay is a probabilistic process; it 
does not occur everywhere simultaneously. In practically all models 
of inflation, false vacuum regions grow due to expansion faster than 
they decay. This means that the total volume of such regions in the universe 
keeps growing without bound. Thus, inflation is a runaway process, which 
stopped in our region, but still continues in other parts of the universe 
(Vilenkin [1983]; Linde [1986]; for a recent review, see Guth [2000]).

Post-inflationary regions like ours form `island universes' in the
inflating sea.  As seen by an observer in the false vacuum region, the
false vacuum is always decaying and so each island universe becomes
larger and larger.  Different parts of the universe spend different
amounts of time in the inflating state.

However, from the point of view of an observer like us in an island
universe, the Big Bang is the time at which inflation ended, and our
notion of time is the time since then.  Our island universe is
infinite, in the sense that the volume of space where the time since
the big bang is the same as that time here goes on forever and so is
infinitely large.  One cannot travel from an island universe to the
inflating sea, because that would require going backward in time, nor
can one travel to another island universe.

The eternally inflating spacetime contains an infinite number of island 
universes. However, since each island universe is itself spatially infinite, 
it is sufficient for our purposes to consider a single island universe.

Remarkably, the entire universe, which contains all these infinite
island universes, may be finite. The apparent contradiction is
resolved due to the fact that the internal notion of time in island
universes is different from the `global' time that one has to use to
describe the entire spacetime.  The volume of the universe at a
particular global time may be finite, but the volume in an island
universe at the time of the Big Bang in that universe (or any later
time) is infinite.

When we discuss the history of a region of our island universe we
include only the history since the Big Bang, i.e., since inflation
ended.  We do not include the time that the region spent in the false
vacuum.  That time has no effect on later events, since the state of a
region at the end of inflation has no dependence on the amount of time
it spent inflating.

\subsection{Every possible history occurs an infinite number of times}

Since the universe is spatially infinite, it can be
subdivided into an infinite number of regions of any given size. Thus
we have an infinite number of regions and only a finite number of
histories that can unfold in them. Since the regions develop
independently, every possible history has a nonzero probability and
will therefore, with probability 1, occur in an infinite number of
regions.  (It is of course possible for an infinite universe
to contain only a finite number of regions with a certain history, but
the probability of that situation is strictly zero, so we will not
consider it.)

Prior to Garriga and Vilenkin ([2001]), a similar argument was given by
Ellis and Brundrit ([1979]), who discussed the implications of the
assumptions that the universe is infinite and approximately
homogeneous. They argued that there should be some regions in such a
universe with histories very similar to that in our region. Our
discussion here goes beyond that of Ellis and Brundrit in two
respects: (i) the spatial infinity of the universe in our picture is a
consequence of the theory of inflation and does not have to be
independently postulated, and (ii) we argued that the number of
distinct histories is finite, which allowed us to conclude that there
should be regions with histories not only similar, but
\textit{identical} to ours (in the sense that those regions contain
qualitative duplicates of every object in our region).

\section{Frequency and Probability}
\label{sec:frequency}

The theory of inflation has surprising consequences for our intuitive 
understanding of \textit{frequency}. On this intuitive understanding, it seems that one should 
be able to obtain exact frequencies by counting up the total quantities of 
certain objects and then doing some simple arithmetic. Thus, suppose that we 
are wondering about the frequency with which planets in the universe contain 
life. Intuitively, it may appear that the exact answer to our question could 
be obtained by counting up all the planets in the universe that contain life 
and then dividing by the total number of planets in the universe.

The theory of inflation shows that this approach is unworkable.  Since
the universe contains infinitely many planets and infinitely many
planets that contain life, no sense can be attached to the notion of a
quotient obtained by dividing the number of planets that contain life
by the total number of planets. Still, there is a certain sense in
which we can speak of the `frequency' with which planets contain
life.  We start out by taking a finite spherical region of space. Then
we can look at the ratio of the number of planets containing life to
the total number of planets in that one finite region. As we increase
the radius of the sphere, this ratio will converge, and the limit will
not depend of the choice of the center of the sphere. The frequency
with which planets in our universe contain life can then be identified
with this limit.

Using this revised definition of frequency, it can be shown that the
frequency of an event is simply equal to its
quantum-mechanical probability. In other words, if quantum
mechanics tells us that some given type of event occurs with
probability $x$, we can infer that that type of event also occurs with
frequency $x$. The quantum-mechanical probability can be
defined in terms of an ensemble of qualitatively identical systems in
the same initial state. The probability that a measurement of some
observable will give a certain result is then simply the fraction
of systems in the ensemble where that result is obtained. With eternal
inflation, there is no need to introduce an imaginary ensemble: for
any system, an infinite ensemble of qualitatively identical systems
exists in each island universe.

We can now introduce the aspect of the theory from which the chief 
philosophical implications will be derived. Although there is an extremely 
small probability that any given region will contain a planet exactly like 
our own---with exactly the same sorts of organisms, exactly the same 
configurations of land and ocean, and so forth---the theory of inflation 
nonetheless permits us to assign a probability of 1 to the 
proposition that there are infinitely many such planets in the universe. 
Moreover, the theory allows us to conclude that the universe contains 
infinitely many planets that diverge from ours in specific ways, with the 
frequency of each type of diverging planet corresponding exactly to its 
probability.

Thus, consider our planet as it was 300 million years ago. Given the exact 
state of our planet at that time, it would be possible (at least in 
principle) to assign quantum-mechanical probabilities to various outcomes. 
There was a certain probability that the planet would eventually come to 
contain mammals, a far smaller probability that the planet would eventually 
come to contain human beings, and so forth. In fact, there was a certain 
probability that the earth would eventually come to contain a human being 
exactly like you, in surroundings exactly like the ones you now inhabit, 
reading a philosophy paper exactly like the one you are reading right now. 
This last probability is extremely small---so small that we could normally 
afford to ignore it. But although the probability is extremely 
small, it is surely above zero.

The theory of inflation now allows us to conclude that, 300 million
years ago, the universe contained infinitely many planets exactly like
our own.  These various planets then underwent various different
histories, with the frequency of each history coming out precisely
equal to its probability. A certain portion contain mammals, a smaller
portion contain humans, and a still smaller portion---almost
unfathomably small, but still nonzero---contain a person exactly like
you.

Our own planet can therefore be seen as one element in an infinite ensemble 
of planets. Indeed, our planet can be seen as an element in a number of 
different infinite ensembles---the ensemble of all planets in the 
universe, the ensemble of all planets that contain intelligent life, the 
ensemble of all planets that are exactly like our own in every respect, and 
so on. In the later sections of the present paper, we argue that a number of 
important philosophical implications can be derived when we regard our 
civilization as an element of one or another of these ensembles.

\section{Inflation Contrasted}

We pause here to compare our theory with three philosophical views that may 
appear (at least on some superficial level) to resemble it. 

Throughout this section, our chief aim is to differentiate the theory of 
inflation from certain philosophical views with which it might be confused. 
At no point will we be arguing that the theory of inflation somehow provides 
evidence in favor of these views. Nor will we claim that it functions as a 
competing theory, such that if the theory of inflation is true, these other 
views must be false. Rather, we claim that the philosophical views are 
directed primarily at questions other than the one that the theory of 
inflation is designed to answer. (Two of the philosophical views are 
concerned primarily with metaphysical questions; the third is concerned 
primarily with ethical questions.) By contrasting the theory of inflation 
with these philosophical ideas, we hope to clarify and further explain 
certain aspects of the theory itself.

\subsection{Modal realism}

First, we should acknowledge that the theory of modal realism, as
formulated by David Lewis ([1986]), appears to yield the very same
conclusion that we have been defending thus far. Lewis is clearly
committed to the view that there are infinitely many regions of any
given size. Moreover, Lewis is committed to the view that every
possible history is realized in at least one region. Indeed
modal realism seems to yield a far stronger conclusion than our own,
since Lewis argues that all \textit{metaphysically possible}
histories are realized, even those that are not physically
possible. It may therefore appear that the theory of inflation is
just a more complicated way of arriving at conclusions that fall
naturally out of Lewis's modal realism.

But this appearance is misleading. Although the theory of inflation and 
modal realism seem to be making similar claims, they are in fact concerned 
with quite different subject matters, and they should therefore be 
regarded as entirely independent. Modal realism is the thesis that all 
possible worlds truly exist. Thus, the modal realist claims that we happen 
to be living in one world (the actual world) but that there are also other 
possible worlds and these other worlds are no less real than our own. By 
contrast, the theory of inflation is a thesis about the \textit{actual} world. The theory 
makes no claims about `other worlds' or `parallel universes.' All of the 
regions posited by the theory are located in the very same spacetime that we 
now inhabit. Thus, when we say that one can assign a probability of 
1 to the proposition that every possible history is realized in infinitely 
many regions, we are making a straightforward physical claim about regions 
of our universe.\endnote{Thus, the theory of inflation should also 
be distinguished from many-world interpretation of quantum mechanics 
(Deutsch [1998]; DeWitt [1970]; Everett [1957]). According to this interpretation, 
the wave function of the universe describes a multitude of disconnected 
universes with all possible histories---a picture reminiscent of the one 
that follows from the theory of inflation. However, the reality of the other 
universes in the many-world theory is still a matter of controversy (see, 
e.g., Brown \& Davies [1993]), whereas the ensemble of regions that we 
discuss in this paper is unquestionably real. (We emphasize that the picture 
of the universe presented here is independent of the interpretation of 
quantum mechanics. If the many-worlds interpretation is adopted, then there 
is an ensemble of disconnected, eternally inflating universes, each having 
an infinite number of regions, where all possible histories unfold. Our 
picture should apply to each of the universes in the ensemble.)} Most of 
these regions are extremely far away, but they are connected to us by 
ordinary spatio-temporal relations, and they all share a common causal 
origin.

For this reason, the theory of eternal inflation is immune to an objection 
that has sometimes been leveled against modal realism. The objection runs 
something like this: `Since events in our own world are supposed to have no 
causal connection to events in other worlds, it seems that we can never 
really learn anything about any world other than our own. Any claim made 
about other possible worlds must be pure speculation, unsupported by the 
usual procedures of scientific inquiry.'

We do not wish to take a position either way about whether or not this is a 
valid objection to modal realism, but we do want to emphasize that the 
theory of inflation is not vulnerable to a parallel objection. The theory of 
inflation is a scientific theory, and it can therefore be supported by 
observational evidence. Of course, someone might argue as follows: `All 
events outside the observable region are, by definition, unobservable. 
Therefore, we cannot gain any knowledge about events outside the observable 
region, and we can never know whether or not every possible history is 
realized in at least one region.' But this argument is without force. First 
of all, it isn't necessarily true that we will never be able to observe 
events outside the observable region. Although we are not \textit{now} able to observe 
such events, we may be able to observe them at some future time. (Indeed, we 
may even be able to travel to parts of the universe that fall outside the 
presently observable region.\endnote{Travel to remote regions may or may 
not be possible, depending on the nature of the dark energy causing the 
accelerated expansion of the universe. If the dark energy density is 
constant, we will not be able to travel beyond the presently observable 
universe. But if the dark energy vanishes over time, then there is no limit 
on how far we can travel.}) More importantly, however, it seems clear that 
we can gain evidence about events in remote regions of the universe without 
ever actually observing those events. Drawing on evidence from the 
observable region, we can construct and test physical theories. These 
theories will then generate predictions about events outside the observable 
region, and insofar as we have reason to believe the theories, we have 
reason to believe the predictions they generate. In other words, even if we 
are never able to make observations concerning events outside the presently 
observable region, our knowledge of the presently observable region may 
permit us to make justifiable inferences concerning events in other parts of 
the universe.

\subsection{Actualism}

Consider now the strong form of determinism according to which nothing can 
possibly happen other than what actually does happen. A proponent of such a 
theory would say, e.g., that if we have actually decided to write this 
paper, we could not possibly have decided not to write the paper, indeed 
that our lives could not have been even slightly different from the way they 
actually are. Following Ayers ([1968]), we refer to this view as 
\textit{actualism}.\endnote{In more recent work, the word `actualism' is normally used to 
refer to the view that only the actual world truly exists (e.g., Adams 
[1981]). Note that we are here using the word in an older sense, such that it 
refers to the view that only actual events are possible.}

It may appear that the actualist arrives ultimately at the very same 
conclusion that we have been defending thus far. After all, it seems that 
actualism and the theory of inflation are simply two different routes to the 
conclusion that everything possible is actual---with the only major 
difference being that actualism claims that surprisingly few things are 
possible whereas the theory of inflation claims that surprisingly many 
things are actual.

But here again, appearances are deceiving. The slogan `Everything possible 
is actual' conceals an important ambiguity, and although this slogan could 
be appropriated with equal justice by either actualism or the theory of 
inflation, it would have very different meanings in these two different 
theoretical contexts.

The actualist asserts that there is only one possible history in any given 
region. By contrast, the theory of inflation does not challenge the 
assumption that, in any given region, there are a variety of distinct 
possible histories. Rather, what the theory asserts is that (with 
probability 1) all of these possible histories will be realized in some 
region of the universe. Thus, although only one of the possible histories 
will be actual in the region that we now inhabit, all possible histories 
will be actual somewhere.

In a certain sense, then, the theory of inflation is the opposite of 
actualism. Daniel Dennett has said that we need to `stave off actualism' 
with `elbow room' that `prevents the possible from shrinking tightly 
around the actual' (Dennett [1984], p. 145, 162) The theory of inflation 
instead posits an infinite amount of space that permits the actual to grow 
to fit the possible.

\subsection{Eternal recurrence}

We turn now to a third philosophical idea that seems to resemble the theory 
of eternal inflation: Nietzsche's doctrine of the eternal recurrence. The 
doctrine is notoriously difficult to interpret, as Nietzsche's published 
works don't include any passages in which he presents it in his own words. 
All interpretations must therefore be based entirely on Nietzsche's 
unpublished notes and on passages from the published works in which 
Nietzsche presents his views through fictional stories.

In these passages, Nietzsche's fictional personae claim that 
everything that has happened in our lives will happen infinitely many times 
in the future (e.g., \textit{Z} III \S2; 
\textit{GS}\S341). Nietzsche scholars disagree about 
how such passages should be interpreted. Some claim that Nietzsche is 
literally advancing a claim about the nature of the universe: namely, that 
every event that we now observe will recur an infinite number of times 
(Danto [1965]). Others say that the doctrine of eternal recurrence should be 
understood not as a literal claim about the nature of the universe but 
rather as a metaphor that we can use to think about our lives. On this 
latter view, the idea is that we ought to live our lives \textit{as though} everything we did 
were going to recur an infinite number of times (Nehamas [1985]). Either way, 
it is clear that Nietzsche meant his doctrine of the eternal recurrence to 
have profound implications for our ordinary decisions. 

Here it might be helpful to consider a more concrete example. Consider a 
novelist who is wondering whether to continue working on his book or just to 
relax for a moment and watch a sit-com on television. And now suppose the 
novelist comes to believe that, whichever action he chooses to perform, that 
action will end up being performed an infinite number of times. It seems 
that the novelist's decision would then acquire an enormous significance, 
what Nietzsche calls `the greatest weight' (\textit{GS} \S341).

Although the theory of inflation seems at least somewhat similar to the 
doctrine of eternal recurrence, it would be wrong to suppose that the theory 
of inflation has the same implications for human life. Like the doctrine of 
eternal recurrence, the theory of inflation says that every action you 
choose to perform will be performed an infinite number of times. But unlike 
the doctrine of eternal recurrence, the theory of inflation also says that 
every possible action you choose \textit{not} to perform will be performed an infinite 
number of times. To get a sense for the force of this claim, consider again 
the novelist facing a decision about how to spend his evening, this time 
assuming that he has come to accept the theory of inflation. The novelist 
will then conclude that there are infinitely many people exactly like him 
and that each member of this infinite ensemble faces a choice between 
working and watching television. However, he will not feel that these other 
people stand to him in any relation of causal dependence.\endnote{Here our 
novelist appears to be faced with a complex problem in decision theory. If 
he chooses to work on his novel, he will be maximizing the expected 
frequency with which his counterparts throughout the universe chose to work 
on their respective novels. (After all, it is highly probable that the 
majority of his counterparts will end up choosing the same option that he 
himself chooses.) But since he cannot actually have any causal impact on 
these counterparts, we will assume that it would be a mistake for him to try 
to maximize the expected frequency with which they perform a particular 
action. In other words, we will presuppose that he ought to act in 
accordance with some version of causal decision theory.} Nor will he believe 
that their choices must necessarily be identical to his own. On the 
contrary, he will reach precisely the opposite conclusion: that no matter 
which option he chooses to select, an infinite number of people exactly like 
him will end up selecting some other option. Thus, he will conclude that, 
even if he chooses to relax and watch television, an infinite number of 
people exactly like him will choose to keep working on the novel.

\section{Ethical Implications}

Since the theory of inflation leads in this way to the opposite conclusion 
from the doctrine of eternal recurrence, one might think that the theory of 
inflation should have the opposite effect on the way people think about 
their lives. Just as the doctrine of eternal recurrence makes every decision 
seem extremely weighty or important, one might think that the theory of 
inflation makes every decision seem insignificant or inconsequential. A 
defender of such a view could say: `We already know (with 
probability 1) that infinitely many good events will occur and that 
infinitely many bad events will occur. We know, for example, that infinitely 
many people exactly like our novelist will finish their work and that 
infinitely many will leave their work unfinished. Nothing that anyone does 
can ever change this. So why should it be a matter of any real concern 
whether some given person happens to choose one option or the other?'

To evaluate this argument, we need to distinguish among a number of 
different ways in which a person might have a deep concern with her own 
decisions. We can then ask, for each of these types of concern, what impact 
the theory of inflation ought to have.

First, let us consider irreducibly \textit{de se} concerns---i.e., concerns that relate 
in some essential way to one's own self (Casta\~neda [1966]; Lewis [1979]). To 
take a simple example, imagine a person who wants to go jogging. Presumably, 
her aim is not that all people of some general type go jogging. Rather, her 
aim is that \textit{she herself} go jogging. To the extent that a person's concerns have this 
\textit{de se} character, they should be relatively unaffected by knowledge of the theory 
of inflation. After all, suppose the person knows full well that there are 
infinitely many people exactly like her, and suppose she knows that, no 
matter what she does, infinitely many of these people will go jogging and 
infinitely many will not go jogging. This knowledge may have little or no 
bearing on her real concern. Her concern is not with what happens to all of 
those other people but with what happens to \textit{her}. She is concerned about whether 
or not \textit{she} ends up going jogging, and the fact that there are infinitely many 
people exactly like her seems not to affect the issue in any way.

Similar remarks apply to those who are concerned with
\textit{particular} objects, events or people. Take the father who
feels a special concern for his own daughter.  Even if he discovers
that remote regions of the universe contain other people who resemble
his daughter in every possible respect, he might find that he cares
far more about his own daughter than he does about any of these other
people (Frankfurt [1999]). Suppose, e.g., that such a man sees his
daughter crying and runs to comfort her. If he accepts the theory of
inflation, he can conclude that the universe contains an infinite
number of events exactly like the one he is now witnessing---an
infinite number of girls exactly like his daughter, all feeling upset
in exactly the same way for exactly the same reason. However, this
conclusion will not lead him to regard his own action as any less
consequential. He will not feel frustrated to learn that he is helping
only one member of an infinite population.  Rather, he will feel that
his own daughter has some special importance---an importance that no
other person can share---and that he is therefore accomplishing
something important by making sure that she receives adequate comfort.

But now suppose we turn to a person who is concerned with the
\textit{total quantity} of something in the world. Such a person might
donate money to the Audubon Society in the hope of increasing the
total quantity of goldfinches. Or, in a more philosophical moment, the
person might think that morality is a matter of increasing the total
quantity of happiness in the universe. Here there really does seem to
be a problem. If there are infinitely many goldfinches in the world,
it seems that one cannot increase their total quantity by donating to
the Audubon Society. Similarly, if there is already an infinite
quantity of happiness, one cannot increase that quantity by engaging
in altruistic activities. (Of course, one can engage in activities
that cause some people to be happier and don't cause any people to be
less happy---but this result is not correctly described as involving a
net increase in any total quantity.) To determine whether or not this
sort of concern should be affected by knowledge of inflation, one has
to ask oneself whether it is truly the total quantity that matters. Is
it necessary that one actually increase the total quantity of
happiness? Or would it be sufficient merely to perform an action that
added some happiness to the world without subtracting any away? Or
would it perhaps be sufficient to increase the total quantity of
happiness \textit{around here} without having any effect at all on the
total quantity in the world as a whole?

Note that the problems that arise here are very different from those
facing the modal realist. The modal realist cannot coherently say that
we ought to try, through our actions, to make reality as a whole
better than it would otherwise have been. Of course, the modal realist
can still use sentences like: `Things would be better if you gave that
money to the Audubon Society than if you gave it to the Nazi Party.'
But such a sentence means only that the possible world in which you
give to the Audubon Society is better than the one in which you give
to the Nazi Party. No real sense can be attached to the notion that
reality as a whole---including all of the possible worlds---would
be better if you did one thing rather than another. Faced with this
difficulty, Lewis ([1986] p.~128) suggests that we give up on the whole
idea of a universalistic ethics. Instead, he argues that we ought to
show a particular concern for the people who occupy our own world.

Eternal inflation is not nearly so radical in its implications. A
believer in eternal inflation can still hold on to the view that the
universe would be better if you did one thing rather than
another. It's just that this notion of the universe's being `better'
cannot be understood in terms of maximizing the total quantity of
happiness. To the extent that one wants to hold on to the underlying
spirit of utilitarianism, one will have to make certain technical
modifications in its `total happiness principle.' Such modifications
have been proposed by Vallentyne and Kagan ([1997]) among others (see
Bostrom [2004]) . We will not be getting into the details here, but
one key implication is that an agent is doing the right thing if he or
she increases the total quantity of happiness of the people in some
finite set and has no effect on anyone else.

Finally, let us consider cases in which a person is specifically concerned 
with \textit{uniqueness}. An art collector may value a particular painting on the grounds that, 
in the whole world, there has never been anything quite like it. A scientist 
may derive a special kind of pride from thoughts like `I am the only person 
ever to have developed this key insight.' An environmentalist may ascribe a 
special importance to a specific herd of animals on the grounds that they 
are the only remaining specimens of their species. Here again, the theory of 
inflation may indicate that something has gone wrong. When claims of 
uniqueness are taken in the most literal sense, the theory of inflation can 
show that they are false. Thus, the art collector is wrong to think that 
there are literally \textit{no} paintings in the entire world exactly like the one she 
now possesses. The truth is that there are infinitely many paintings exactly 
like hers; it's just that they are so far away that she will never be able 
to observe them. The important question, then, is whether it really matters 
that a particular object or event be literally \textit{unique}. Does it really matter, for 
example, that the painting be literally the only one of its kind in the 
entire universe? Or is it sufficient that the painting be the only one of 
its kind within a 10$^{100}$ parsec radius?

This sort of question becomes especially pressing when applied to the
concern we feel about the continuing existence of our own
civilization. The theory of inflation tells us that the universe
contains an infinite number of civilizations exactly like ours. Thus,
even if our own civilization is entirely destroyed over the course of
the next century, the theory tells us that an infinite number of other
civilizations exactly like ours will continue to exist. Does the
theory therefore give us a reason to feel less concerned about nuclear
wars, asteroid collisions and other events that might destroy our
civilization? Here again, the answer will depend on why exactly we
were concerned about the possibility of this destruction in the first
place. If we were concerned because we valued particular people or
particular institutions that now inhabit the earth, then the theory
should have no effect on our feelings. But if we were concerned
because we felt that our civilization was somehow \textit{unique}---so
that if our civilization were destroyed, the universe would no longer
contain anything even remotely like the presently-existing human
race---then the theory tells us that our concern was based on a false
assumption.

Presumably, the concern that we actually feel is based on a complex 
combination of different beliefs, desires and emotions. Some of these should 
be affected by the theory of inflation; others should not. It therefore 
remains to be seen whether the theory should have any substantial impact on 
our overall attitude toward the continuing existence of our civilization.

\section{Universal Doomsday}

As discussed above, the theory of inflation implies that we are part of an 
infinitely large `island universe' that contains an infinite number of 
civilizations. According to the anthropic principle (codified, for example, 
as the `self-sampling assumption'; Bostrom [2002]) we should reason as if we 
were randomly selected from all the individuals in all those civilizations. 
Thus our expectation of finding ourselves in any particular circumstances is 
proportional to the number of observers in those circumstances. We now want 
to ask whether it is possible to use information about our own circumstances 
to make inferences about the average lifetimes of civilizations in our 
universe. 

First, it is clear that there is some nonzero probability for a
civilization to survive early threats to its existence (nuclear war,
asteroid impact, etc.). Such a civilization might go on to spread
across its galaxy. It could endure for millions of years and contain a
huge number of individuals. We will refer to such civilizations as
\textit{long-lived}. On the other hand, some civilizations will
succumb to existential threats and so be \textit{short-lived}. What
will be the fraction of each?

Unless the fraction of long-lived civilizations is tiny, nearly all 
individuals will belong to them, and furthermore will live late in their 
civilizations when most of the individuals live. That, however, is not the 
circumstance in which we find ourselves. Instead, we find that we live 
either in a short-lived civilization or very early in a long-lived one. 
While we do not have a clear idea of how long to expect civilizations to 
last, when we take into account our circumstances, we should clearly update 
our ideas in favor of a much larger chance for civilizations to be 
short-lived (Carter unpublished; Leslie [1996] p.~231; see also Olum [2004]).

Let $f(S)$ be the frequency (as defined in
\S\ref{sec:frequency}) with which civilizations are short-lived, and $f(L)
= 1-f(S)$ the frequency with which civilizations are long-lived.  We
don't know these frequencies, so let $f_{\rm prior}$ denote our best
understanding of the frequencies before we take into account our
circumstances.  Now suppose you learn that you have birth rank $N$,
i.e., you are the $N$th human to be born.  You should
update your frequencies via Bayes's Rule, to get
\begin{equation}\label{eqn:Bayes} 
f_N (S) ={P (N|S) f_{\rm prior} (S)\over P (N|S) f_{\rm prior} (S)
+P (N|L) f_{\rm prior} (L)}\quad,
\end{equation}
where $P (N|a)$ is the chance that you would have birth rank $N$ given
that we are in a long- or short-lived civilization, and $a$ is $L$ or
$S$ accordingly.  You could be any individual in the civilization, so
\begin{equation}
P(N|a) = \cases{1/N_a& $N \le N_a$\cr0 & $N > N_a$}\,,
\end{equation}
where $N_a$ is the number of individuals in each
civilization.  Since $N_L\gg N_S$, $P (N|L)\ll P (N|S)$, and thus $f_N
(S)$ is nearly 1 unless $f_{\rm prior}(S)$ is extremely small.  Unless we
previously thought that long-lived civilizations were much more
likely, we should now think that almost all civilizations will be
short-lived---a sort of `universal doomsday'.

The `universal doomsday' argument that we advance here should be
carefully distinguished from the classic doomsday argument (Carter
unpublished; Gott [1993]; Leslie [1989], [1996]; Nielsen [1989]). The classic
doomsday argument was an attempt to show that our present
circumstances give us some reason to believe that \textit{our own
particular} civilization will soon come to an end. The argument
advanced here is quite different. We make no specific claims
regarding the longevity of any particular civilization. Rather, we say
that our present circumstances give us reason to reach a general
conclusion about our universe: namely, that long-lived civilizations
are extremely infrequent in our universe as a whole.

Moreover, as we now proceed to argue, the theory of inflation gives us 
reason to reject the particular doomsday argument, accepting only the 
universal doomsday argument. Thus, the doomsday argument has nothing to say 
specifically about our own civilization as distinct from others. Instead it 
tells us about the general longevity of civilizations sufficiently similar 
to ours to be included in the same reference class---although, of course, 
what we learn about civilizations in general, we should also apply to 
ourselves.

\subsection{Application to our civilization in particular} 

Traditionally, the doomsday argument has been applied to the future of our 
own particular civilization.  We let $P_{\rm prior} (a)$, with $a =
S$ or $L$, be our subjective probabilities that our own civilization will be
short-lived or long-lived, before we take into account
our birth rank.  We then update our probabilities as
\begin{equation}\label{eqn:PBayes}
P (S|N) ={P (N|S) P_{\rm prior} (S)\over P (N|S) P_{\rm prior} (S)
+P (N|L) P_{\rm prior} (L)}
\end{equation}
Thus unless $P_{\rm prior}(S)$ is infinitesimal, we will find that $P (S|N)$
is nearly 1, just as above.

That analysis, however, neglects the possibility that we could have
been in any other civilization.  There is some controversy about which
individuals should be included in the reference class among which we
should expect to be typical, but it should be clear that we must at
least include all observers subjectively indistinguishable from
ourselves (Bostrom [2002]). However, the theory of inflation implies
that there are infinitely many such observers, belonging to
civilizations with every possible lifespan.

Now we should reason as though we were chosen randomly among humans in
the various human civilizations in our universe, where again the
random choice is to be understood as the limit of random choices in
successively larger finite regions, as in \S\ref{sec:frequency}.
Then, before we make the probability shift above, we should
realize that the chance that our civilization is large is not just the
frequency of civilizations that are large, but must be corrected for
the increased chance to be in a long-lived civilization because it has
more individuals.  Taking that into account, we find that the
probability for our civilization to be small, before taking into
account birth rank, is
\begin{equation}\label{eqn:goodprior}
P_{\rm prior} (S) ={N_S f(S)\over N_S f(S) +N_L f(L)}
\end{equation}
and similarly for $P_{\rm prior}(L)$.  To take our birth rank into account, we
use (\ref{eqn:goodprior}) in (\ref{eqn:PBayes}) to get
\begin{equation}
P (S|N) =f (S)\,.
\end{equation}

This effect exactly cancels out the impact of the particular 
doomsday argument, leaving us with the conclusion that our chances that we 
are now in a long-lived or a short-lived civilization are just proportional 
to the prevalence of such civilizations (Bostrom [2002]; Dieks [1992]; Olum 
[2002]). Thus, if the theory of inflation is correct, the doomsday argument 
has nothing to say about the longevity of our specific civilization, but 
only about the general longevity of civilizations sufficiently similar to 
ours to be included in the same reference class.

At this point, one might object that similar considerations could be
used to defeat the universal doomsday argument. Thus, one might
suggest that we are typical not merely among all those individuals in
our universe, but rather among all those individuals who might exist
according to alternative theories of the universe, if we don't know
which theory is correct. If the universe developed in some
probabilistic way before the beginning of inflation, so that early
chance events affected all regions together, then one can consider
also the possible observers who might exist as a result of all
different early developments. Including all such possible observers in
the reference class is equivalent to accepting the self-indication
assumption (SIA) (Bostrom [2002]), first introduced by Dieks ([1992]),
which states that the chance that you would exist at all is greater in
a universe which contains more observers. If one accepts SIA, then a
universe with long-lived civilizations is more likely because of the
greater number of individuals that it contains, and that effect
cancels all forms of doomsday argument. SIA is
controversial, however, most notably because of the `presumptuous
philosopher' example (Bostrom [2002]).  We will not address that
controversy here\endnote{For recent discussions of arguments for and
against SIA, see Olum ([2002]) and Bostrom and Cirkovic ([2003]).}, but
for the purposes of the present paper will consider the situation
where one does not include SIA.

\subsection{Universal vs. particular dooms}

Some effects which might shorten the life expectancy of our civilization 
apply only to ours specifically, while others shorten the general life 
expectancy of all civilizations. For example, suppose that we are concerned 
with the earth being hit by an asteroid. The chance of such a collision, in 
the next century say, is a function of the number of asteroids in the solar 
system and the chance that any given asteroid is on a course which will hit 
the earth during that period. 

Now a specific asteroid which happens to be on a collision course with us is 
a `particular doom' that affects only us. The fact that the asteroid has, 
by chance, the doomsday orbit says nothing about other asteroids in other 
solar systems like ours. The particular orbit of the asteroid is unrelated 
to the distribution of civilizations that will or will not be destroyed by 
asteroids. Given the theory of inflation, there is thus no reason to believe 
that such orbits are more likely than one would first think. 

On the other hand, the total number of asteroids could well be determined by 
some universal process of solar system formation and most solar systems like 
ours would have similar numbers of asteroids. Therefore if the (incompletely 
known) process that produces asteroid belts turns out to produce an 
especially large number of asteroids, the lifetimes of all civilizations 
would be on average shortened. Large numbers of asteroids are a `universal 
doom' that (statistically) affects all civilizations, and thus the doomsday 
argument makes them more likely. 

\subsection{Practical applications}

The doomsday argument has practical applications.  If you care about
the future of the human race you might want to reduce the probability
of possible causes of extinction that you might have some control
over.  The degree to which you should be concerned with such possible
causes depends on how likely they are, and if you accept the doomsday
argument you should think the possible dooms more likely than you
would otherwise have thought.

The argument presented here should change your calculation.  You
should not think a process is more likely if it affects us alone, but
you should think it more likely if it affects civilizations everywhere
in the universe.  To continue the above example, you should be more
concerned that a large number of asteroids have not yet been detected
than about the particular orbit of each one. You should not worry
especially about the chance that some specific nearby star will become
a supernova, but more about the chance that supernovas are more deadly
to nearby life than we believe. Many other examples are possible.

\section{Concluding Remarks}

Since at least the time of Copernicus, physicists have been casting doubt on 
the na\"\i ve view that our planet plays some unique and special role in 
the universe. First it became clear that our planet was not the center of 
the cosmos---that the planet Earth was just one of the planets in our solar 
system. Then we gradually accumulated evidence for the view that our solar 
system was itself just one of the many such systems in the universe. These 
theoretical advances contributed to a growing sense that our civilization 
plays no special role in the cosmic drama, that it is just one tiny speck in 
a vast universe. Thus, a series of scientific discoveries led to a series of 
philosophical problems---problems about the significance of human life, 
about our role in the divine plan, and so forth. 

But although scientific discoveries have done a great deal to threaten
our na\"\i ve worldview, they did appear to leave us with one way of
holding on to our intuitive sense that there was something special and
unique about the planet earth. We knew that our planet was just one of
the many planets in the universe, but we could nonetheless hold on to
the idea that it was the only planet that had certain distinctive
properties---probably the only planet with anything remotely like a
human being, certainly the only one with all the art forms, cultural
traditions and political institutions that we most associate with life
on earth. The theory of inflation now shows us that even this last
claim to uniqueness was, in fact, illusory. Even we ourselves are not
unique; in that sense, as Alan Guth has said, the theory shows that we
do not even have `a unique copyright on our own identities' (Quoted in
Martin [2001]). This new theoretical advance casts up a set of new
philosophical questions; we have tried to begin the exploration of
those questions here.

\section*{Acknowledgments}

We are grateful to Nick Bostrom, Brandon Carter, Daniel Dennett, Jaume
Garriga and two anonymous referees for useful discussions and
comments. The work of K.D.O. and A.V. was supported in part by the
National Science Foundation.



\begin{flushright}
\textit{Philosophy Department, Princeton University}\\
\textit{Princeton, NJ 08544-1006}\\
jknobe@Princeton.edu
\end{flushright}
\begin{flushright}
\textit{Department of Physics and Astronomy, Tufts University}\\
\textit{Medford, MA 02155}\\
kdo@cosmos.phy.tufts.edu
\end{flushright}
\begin{flushright}
\textit{Department of Physics and Astronomy, Tufts University}\\
\textit{Medford, MA 02155}\\
vilenkin@cosmos.phy.tufts.edu
\end{flushright}

\newpage 
\section*{References}
\parindent 0pt
\parskip 2pt
Adams, R. [1981]: `Actualism and Thisness', \textit{Synthese}, \textbf{49}, pp. 3-41. 

Ayers, M. [1968]: \textit{The Refutation of Determinism}, London: Methuen.

Bostrom, N. [2002]: \textit{Anthropic Bias: Observation Selection Effects}, New York: Routledge.

Bostrom, N. [2004]: `Infinite Ethics',\newline
http://www.nickbostrom.com/ethics/infinite.pdf

Bostrom, N. and M. Cirkovic [2003]: `The Doomsday Argument and the 
Self-Sampling Assumption: Reply to Olum', \textit{Philosophical Quarterly}, \textbf{53}, pp. 83-91.

Brown, J. and P. C. W. Davies (\textit{eds.}) [1993]: \textit{The Ghost in the Atom} (Cambridge: Cambridge University 
Press).

Casta\~{n}eda, H.-N. [1966]: `He: A Study in the Logic of 
Self-Consciousness', \textit{Ratio}, \textbf{8}, pp. 130-57.

Danto, A. [1965]: \textit{Nietzsche as Philosopher}, New York: Macmillan.

Dennett, D. [1984]: \textit{Elbow Room: The Varieties of Free Will Worth Wanting}, Cambridge, Mass.: MIT Press.

Deutsch, D. [1998]: \textit{The Fabric of Reality}, London: Penguin Books.

DeWitt, B. [1970]: `Quantum Mechanics and Reality', \textit{Physics Today}, \textbf{23}, pp. 30-40.

Dieks, D. [1992]: `Doomsday --- or: The Dangers of Statistics', 
\textit{Philosophical Quarterly}, \textbf{42}, pp. 78-84.

Ellis, G. F. R. and G. B. Brundrit [1979]: `Life in the Infinite 
Universe', \textit{Quarterly Journal of the Royal Astronomical Society}, \textbf{20}, pp. 37-41.

Everett, H. [1957]: `Relative-State Formulation of Quantum Mechanics', 
\textit{Reviews of Modern Physics}, \textbf{29}, pp. 454-62.

Frankfurt, H. [1999]: `On Caring', In his \textit{Necessity, Volition, and Love} (Cambridge: Cambridge University Press). 

Garriga, J. and A. Vilenkin [2001]: `Many worlds in one', 
\textit{Physical Review D}, \textbf{64}, 043511.

Gell-Mann, M. and J. B. Hartle [1993]:
`Classical Equations For Quantum Systems,'
\textit{Physical Review D}, \textbf{47}, pp. 3345-82

Gott, J. R. III [1993]: `Implications of the Copernican Principle for 
our Future Prospects', \textit{Nature}, \textbf{363}, pp. 315-9.

Guth, A. [1981]: `The Inflationary Universe: A Possible Solution to the 
Horizon and Flatness Problems', \textit{Physical Review D}, \textbf{23}, pp. 347-56.

Guth, A. [2000]: `Inflation and Eternal Inflation', \textit{Physics Reports}, \textbf{333}, pp. 555-74.

Leslie, J. [1989]: `Risking the World's End', \textit{Bulletin of the Canadian Nuclear Society}, \textbf{10}, pp. 10-5.

Leslie, J. [1996]: \textit{The end of the world}, London: Routledge.

Lewis, D. [1979]: `Attitudes \textit{De Dicto} and \textit{De Se}', \textit{The Philosophical Review}, \textbf{88}, pp. 513-43.

Lewis, D. [1986]: \textit{On the plurality of worlds}, Oxford: Blackwell.

Linde, A. [1986]: `Eternally Existing Self-Reproducing Chaotic 
Inflationary Universe', \textit{Physics Letters}, \textbf{B175}, pp. 395-400.

Martin, M. [2001]: `Other Worlds May Surround Us, Physicists Claim', 
United Press International,\newline 
http://www.sciencenewsweek.com/articles/worlds.htm. 

Nehamas, A. [1985]: \textit{Nietzsche: Life as literature}, Cambridge, Mass.: Harvard University Press.

Nielsen, H. B. [1989]: `Random Dynamics and Relations Between the 
Number of Fermion Generations and the Fine Structure Constants', \textit{Acta Physica Polonica B}, \textbf{20}, pp. 427-68.

Nietzsche, F. [1969]: \textit{Thus spoke Zarathustra} [Cited as \textit{Z}.] Trans. R. J. Hollingdale (London: 
Penguin Books).

Nietzsche, F. [1974]: \textit{The Gay Science} [Cited as \textit{GS}.] Trans. W. Kaufmann. New York: Random House.

Nietzsche, F. [1968]. \textit{The Will to Power} [Cited as \textit{WP}.] Trans. W. Kaufmann and R. J. 
Hollingdale, New York: Random House.

Olum, K. D. [2002]: `The doomsday argument and the number of possible 
observers', \textit{Philosophical Quarterly}, \textbf{52}, pp. 164-84.

Olum, K. D. [2004]: `Conflict between anthropic reasoning and
observation', \textit{Analysis}, \textbf{64}, pp. 1-8.

Richards, T. [1975]: `The Worlds of David Lewis', \textit{Australasian Journal of Philosophy}, \textbf{53}, pp. 105-18.

Smolin, L. [2001]: Talk given at the conference \textit{Anthropic arguments in fundamental physics}, Cambridge.

Vallentyne, P. and S. Kagan [1997]: `Infinite Value and Finitely 
Additive Value Theory', \textit{The Journal of Philosophy}, \textbf{94}, pp. 5-26.

Vilenkin, A. [1983]: `The birth of inflationary universes', \textit{Physical Review D27}, pp. 
2848-55.

Vilenkin, A. [1998]: `Unambiguous probabilities in an eternally 
inflating universe', \textit{Physical Review Letters 81}, pp. 5501-4.

\end{document}